\documentclass{jfm}
\usepackage{graphicx}
\usepackage{natbib}
\usepackage{amssymb}
\usepackage{amsmath}
\usepackage[usenames]{color}
\usepackage[normalem]{ulem}
\renewcommand{\sout}[1]{} 
\usepackage{subcaption}

\newcommand{\MDG}[1]{\textcolor{blue}{*** #1 ***}}

\newcommand{\AS}[1]{\textcolor{black}{#1}}
\newcommand{\AScomment}[1]{\textcolor{black}{#1}}
\newcommand{\MDGrevise}[1]{\textcolor{black}{#1}}
\newcommand{\ASrevise}[1]{\textcolor{black}{#1}}
\newcommand{\mbf}[1]{\boldsymbol{#1}}
\newcommand{\Wi}{\textit{Wi}}

\title{Self-sustained elastoinertial Tollmien-Schlichting waves}

\author{Ashwin Shekar\aff{1}, Ryan M. McMullen\aff{2}, Beverley J. McKeon\aff{2} \and  Michael D.~Graham\aff{1}\corresp{\email{mdgraham@wisc.edu}}}
\affiliation{\aff{1}Department of Chemical and Biological Engineering, University of Wisconsin-Madison, Madison, WI 53706, USA
\aff{2}Graduate Aerospace Laboratories,
	California Institute of Technology,
	Pasadena CA 91125, USA}

\begin{document}

\maketitle   

\begin{abstract}
        Direct simulations of two-dimensional plane channel flow of a viscoelastic fluid at Reynolds number $\Rey=3000$ reveal the existence of a family of attractors whose structure closely resembles the linear Tollmien-Schlichting (TS) mode, and in particular exhibits strongly localized stress fluctuations at the critical layer position of the TS mode. At the parameter values chosen, this solution branch is not connected to the nonlinear TS solution branch found for Newtonian flow, and thus represents a \sout{new} solution family that is nonlinearly self-sustained by viscoelasticity. The ratio between stress and velocity fluctuations is in quantitative agreement for the attractor and the linear TS mode, and increases strongly with Weissenberg number, $\Wi$. For the latter, there is a transition in the scaling of this ratio as $\Wi$ increases, and the $\Wi$ at which the nonlinear solution family comes into existence is just above this transition. Finally, evidence indicates that this branch is connected through an unstable solution branch to two-dimensional elastoinertial turbulence (EIT). These results suggest that, in the parameter range considered here, the bypass transition leading to EIT is mediated by nonlinear amplification and self-sustenance of perturbations that excite the Tollmien-Schlichting mode. 

\end{abstract}

\section{Introduction}


Adding minute quantities (parts per million) of long-chain polymer additives can dramatically change the turbulent flow of Newtonian fluids, the most significant consequence being the considerable drop in friction factor, which is commonly referred to as the Toms effect \citep{VIRK:1975vb,White:2008hs,Graham:2014uj}. Accompanying this overall change is a structural change to the flow. At high levels of viscoelasticity, \cite{Samanta:2013el} and \cite{Dubief:2013hh} have shown that trains of weak spanwise-oriented flow structures with inclined sheets of polymer stretch dominate the flow, denoting this regime as elastoinertial turbulence (EIT). {These sheets of polymer stretch correspond with a layer near each wall of  localized spanwise vortex motion.} In further contrast to the 3D structures that sustain Newtonian turbulence, \cite{Sid:2018gh} demonstrated that EIT is fundamentally 2D in nature by showing that the structures sustaining 2D EIT in channel flow simulations are similar to those in 3D. {In a computational study of EIT in pipe flow, \cite{lopez2019dynamics} also observe nearly two-dimensional spanwise vortices in the near-wall region, indicating the structural similarities between EIT in channels and pipes.}

\cite{Choueiri:2018it} experimentally studied the path to EIT in pipe flow by varying  polymer concentration at fixed 
Reynolds number, $\Rey$. For sufficiently low $\Rey$, they observed an initial drop in friction factor (i.e.~modification of Newtonian turbulence) as concentration increased, followed by re-laminarization and eventually by a reentrant transition to EIT, where the flow had  very different structure from Newtonian turbulence. These observations point at two distinct types of turbulence in dilute polymer solutions --  one that is suppressed by viscoelasticity (Newtonian turbulence) and one that is promoted (EIT).

\cite{Shekar:2019hq} corroborated this observation of a reentrant transition to EIT in simulations of channel flow with increasing Weissenberg number, $\Wi$, the ratio between the polymer relaxation time scale and the shear time scale. They further showed that close to its inception, EIT exhibits localized polymer stress fluctuations that bear strong resemblance to critical layer structures predicted by linear analyses, i.e., sheetlike fluctuations localized at wall-normal locations where the disturbance wavespeed equals the base flow velocity. 
In particular, they demonstrated that the fluctuation structure corresponding to the dominant spectral content strongly resembles the viscoelastic extension of the linear Tollmien-Schlichting (TS) wave. This is perhaps a surprising result, as the flow in the parameter regime considered is linearly stable, and in Newtonian turbulence, the TS mode plays a very limited role. 
Some light is shed on this issue through resolvent analysis, i.e., determination of the response of the linearized dynamics to harmonic-in-time disturbances, which shows that the linear TS mode becomes highly amplified in the presence of viscoelasticity. This strong amplification implies that even very weak disturbances may be sufficient to trigger the nonlinear effects necessary to sustain EIT.

We note that similar structures have been observed by other researchers in different contexts. 
\cite{Page:2015es} 
analyzed the evolution of vortical perturbations in 2D viscoelastic simple shear flow. Their analysis reveals a viscoelastic analogue of the Newtonian Orr mechanism. This ``reverse-Orr'' mechanism generates tilted sheets of polymer stress fluctuations resembling those seen at EIT and thus may play some role in this phenomenon. 

Because prior work on elastoinertial turbulence reveals structures similar to those seen in the linear Tollmien-Schliching mode, the present work focuses on Tollmien-Schlichting waves, but in the fully nonlinear context of self-sustained solutions in the channel flow geometry.  (In the parameter regime here, the laminar flow is always linearly stable.) After introducing the formulation and computational methods, we show how the Newtonian nonlinear TS wave branch is modified by viscoelasticity, resulting in its disappearance as $\Wi$ increases. At still higher $\Wi$, however, we demonstrate the onset of a new, viscoelasticity-driven, nonlinear solution branch that strongly resembles the linear Tollmien-Schlichting mode, and illustrate how it is related to the TS mode of linear theory and to elastoinertial turbulence.


\section{Formulation}

This study focuses on two-dimensional pressure-driven channel flow
with constant mass flux. The $x$ and $y$ axes are aligned with the streamwise and wall-normal directions, respectively. 
Lengths are scaled by the half channel height $l$, so the dimensionless channel height $L_y=2$. The domain is periodic in $x$ with length $L_x$. 
Velocity $\mbf{v}$ is scaled with the Newtonian laminar centerline velocity $U$; time $t$ with $l/U$, and pressure $p$ with $\rho U^2$, where $\rho$ is the fluid density. The polymer stress tensor $\mbf{\tau}_p$ is related to the polymer conformation tensor $\mbf{\alpha}$ 
 through the FENE-P constitutive relation, which models each polymer molecule as a pair of beads connected by a nonlinear spring with maximum extensibility $b$.
 
We solve the momentum, continuity and FENE-P equations:

\begin{align}
        \label{Eq_ns_momentum}
                \frac{\partial \mbf{v}}{\partial t} +
                \mbf{v} \cdot \mbf{\nabla v} = -
                \mbf{\nabla}p + \frac{\beta}{Re} \nabla^{2}\mbf{v} +
                \frac{\left(1 -\beta\right)}{Re \Wi_{}}\left(\mbf{\nabla} \cdot
                \mbf{\tau}_{\mathrm{p}}\right), \\
                  \mbf{\nabla} \cdot \mbf{v} = 0,   \\
        	        \mbf{\tau}_p = \frac{\mbf{\alpha}}{1-\frac{\mathrm{tr}(\mbf{\alpha})}{b}} - \mbf{I}, \\             
	    			\frac{\partial \mbf{\alpha}}{\partial t} +         
        			\mbf{v} \cdot \mbf{\nabla \alpha} -
        			\mbf{\alpha} \cdot \mbf{\nabla v} - 
        			\left( \mbf{\alpha} \cdot \mbf{\nabla v} \right)^{\mathrm{T}}
			    = \frac{-1}{\Wi_{}} \mbf{\tau}_p.
\end{align}
Here $Re = \rho U l / (\eta_{\mathrm{s}} + \eta_{\mathrm{p}})$, where
$\eta_s$ and $\eta_p$ are the solvent and polymer contributions to the zero-shear rate viscosity.
The viscosity ratio $\beta = \eta_{\mathrm{s}} / (\eta_{\mathrm{s}} + \eta_{\mathrm{p}})$.  We fix $\beta=0.97$ and $b=6400$. Since $1-\beta$ is proportional to polymer concentration and $b$ to the number of monomer units, this parameter set corresponds to a dilute solution of a high molecular weight polymer.
The Weissenberg number $\Wi = \lambda U/l$, where $\lambda$ is the polymer relaxation time. 

For the nonlinear direct {numerical} simulations {(DNS)} described below, a finite difference scheme and a fractional time step method are adopted for integrating the Navier-Stokes equation. Second-order Adams-Bashforth and Crank-Nicolson methods are used for convection and diffusion terms, respectively. The FENE-P equation is discretized using a high resolution central difference scheme \citep{kurganov2000new,Vaithianathan:2006dy, Dallas:2010gu} 
\sout{No artificial diffusion is applied.} 
\AS{that guarantees positive definiteness of the polymer conformation tensor without the need for any artificial diffusion. In any case, the nonlinear solution branch on which we focus in this manuscript displays weak fluctuations far from the limit of positive definiteness even at the highest $\Wi$ of existence.}
\sout{Resolution tests were performed to ensure convergence of statistics.} A typical resolution for the following results is $(N_{x},N_{y})$ = ${(79, 402)}$. \AS{\sout{R1a, R3c - }This resolution used was based on mesh convergence results at $\Wi = 45$. When the resolution was increased to $(N_{x},N_{y})$ = ${(131, 602)}$, the mean polymer stretch deviations from the laminar base state change by less than 1 percent.}
 
We also consider the linearized evolution of infinitesimal perturbations to the laminar state with given streamwise wavenumber $k$. Two approaches are used. The first is classical linear stability analysis, in which solutions of the form 
$\mbf{\hat{\phi}}(y)\exp\left[ik(x - ct)\right]$ are sought, resulting in an eigenvalue problem for the complex wavespeed $c$ at a given $k$. In the present study, $~\hat{}~$~always indicates deviation from the laminar state. 
If all $c_i<0$, which is the case for all conditions considered in the present study, the flow is linearly stable. \ASrevise{A linearized version of the DNS code was also developed using the numerical schemes described above. Results were validated against linear stability analysis and agreement to three significant digits was obtained for the value of $c$ for the viscoelastic TS mode at the parameters of interest.} 

The second linear approach used here determines the linear response of the laminar flow to external forcing with given wavenumber $k$ and frequency $\omega$ using the resolvent operator of the linearized equations \citep{Schmid07,Mckeon:2010ep}. The norm used in the resolvent calculations is 
\begin{equation}
\lVert \mbf{\hat{\phi}} \rVert^2_{\mbf{A}} = \int_{-1}^{1} \left[ \mbf{\hat{v}}^*\mbf{\hat{v}} + \mathrm{tr}\!\left( \mbf{A}^{-1}\mbf{\hat{\alpha}}^*\mbf{A}^{-1}\mbf{\hat{\alpha}} \right) \right] \mathrm{d}y,
\end{equation}
where $\mbf{A}$ is the conformation tensor in the laminar state. The second term provides a measure of the conformation tensor perturbation magnitude that is motivated by the non-Euclidean geometry of positive-definite tensors~\citep{hameduddin2019}.  
For both the linear stability and resolvent analyses, the equations are discretized with a Chebyshev pseudospectral method using 401 Chebyshev polynomials. {This number was arrived at by ensuring convergence of the TS eigenvalue.}



\section{Results and discussion}

\subsection{Origin of Newtonian and viscoelastic nonlinear Tollmien-Schlichting attractors}

\begin{figure}
	\begin{center}
		\begin{subfigure}{0.45\textwidth}
			\includegraphics[scale=0.1]{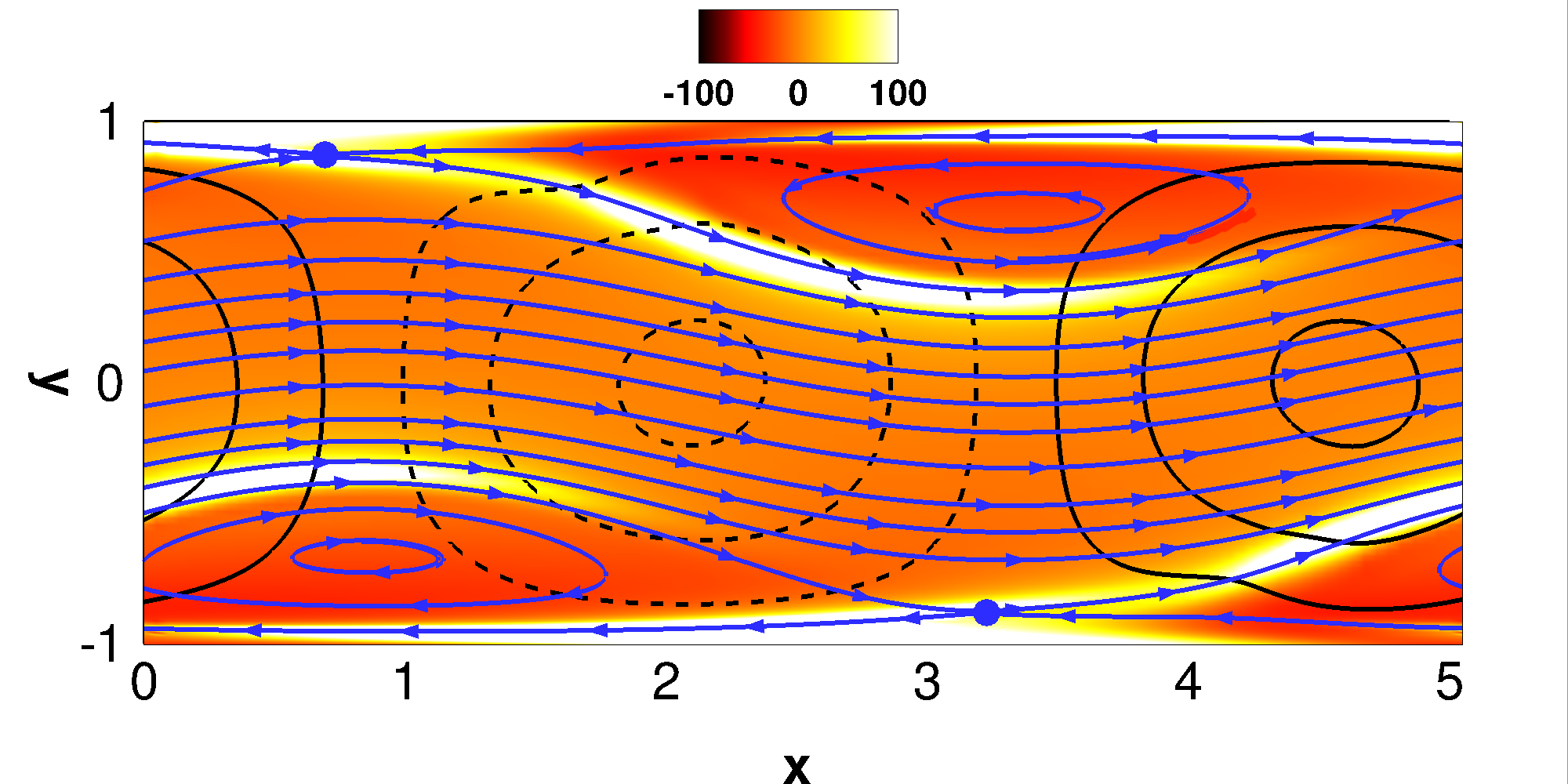} 
			\label{fig:NL_TSW_Re3000_Wi3}
		\end{subfigure}
		\caption[]{Structure of NNTSA at $\Rey = 3000, \Wi = 3$. Streamlines (blue) in a reference frame moving at the wave speed $c = 0.39$ are superimposed on color contours of $\hat{\alpha}_{xx}$ where~$\hat{}$ denotes deviations from the laminar state. Blue dots indicate the locations of hyperbolic stagnation points (in the traveling frame). Black contour lines of $\hat{v}$ are also shown. For $\hat{v}$, dashed = negative and solid = positive.}
		\label{fig:Streamline_7}	     
	\end{center}
\end{figure}

In Newtonian flow, a family of nonlinear Tollmien-Schlichting waves bifurcates subcritically from the laminar branch at $\Rey\approx 5772$ with $L_x\approx 2\pi/1.02\approx 6.15$.  The lower limit of the parameter regime for which this solution family exists is $\Rey\approx 2800, L_x\approx 2\pi/1.3\approx 4.83$ \citep{Jimenez:1990cn}. Furthermore, in prior work on elastoinertial turbulence \citep{Shekar:2019hq}, as well as more recent simulations in long two-dimensional domains, a strong peak in the spatial spectrum is found at $L_x\approx 5$.
Based on these observations, all of the results presented in this study will be at $\Rey=3000, L_x=5$. \AS{(At $L_x = 5$, Newtonian channel flow is linearly stable at all $\Rey$.)}
In the Newtonian limit at these parameters, there are upper and lower branch solutions (which merge in a saddle-node bifurcation as $\Rey$ is lowered); the upper branch traveling wave solution is linearly stable with respect to two-dimensional perturbations and is thus easily computed via DNS. We call this solution branch, including its viscoelastic extension, the Newtonian Nonlinear Tollmien-Schlichting attractor (NNTSA). (The word ``attractor" is chosen rather than ``wave'' because, depending on parameters one can observe a pure traveling wave state or one with periodic or nonperiodic modulations.)   

On increasing $Wi$, the self-sustained nonlinear viscoelastic TS wave at $Re=3000$ develops \sout{a structure of}sheets of high polymer stretch 
\sout{bearing resemblance to }\sout{\MDGrevise{resembling} near wall structures seen at EIT.}\AS{\sout{R1c - }that start out from near the wall. These observations are evidence of the capability of nonlinear TS critical layer mechanisms in generating sheets of polymer stretch\sout{ which could play a role at EIT}.}
Figure \ref{fig:Streamline_7} illustrates this {point} with a snapshot of $\hat{\alpha}_{xx}$ on the NNTSA branch at $Wi=3$. 
The sheets originate in the nonlinear Kelvin cat's eye kinematics of TS waves at finite amplitude, as detailed in \cite{Shekar:2019hq}. \AS{The NNTSA continues to display  wall normal velocity fluctuations that extend across the channel centerline -- a signature of TS kinematics.} Some of the observations made in \cite{Shekar:2019hq} are repeated here {for completeness}, as they form the background for the new results of the present study.

\begin{figure}
	\begin{center}
		\begin{subfigure}{0.45\textwidth}
			\includegraphics[scale=0.28]{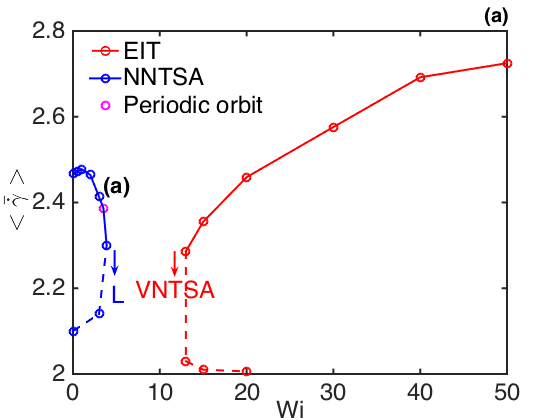}
			\label{fig:EIT_BFD}
		\end{subfigure}
		\begin{subfigure}{0.45\textwidth}
			\includegraphics[scale=0.28]{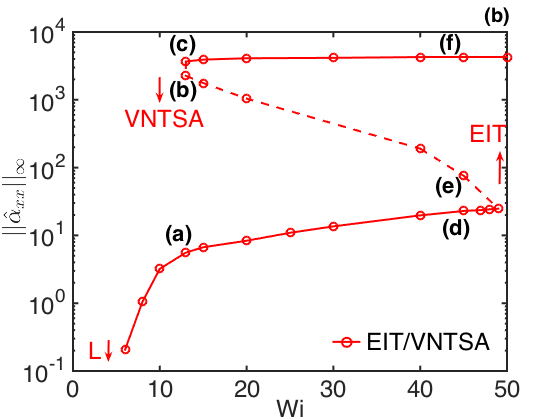} 
			\label{fig:BFD_Comb}
		\end{subfigure}
		\caption[]{(a) Bifurcation diagram showing the evolution of the space and time-averaged wall shear rate  with $Wi$ for the 2D nonlinear NNTSA and EIT branches at $Re = 3000, L_x=5$. \AS{Point (a) corresponds to the structure shown in Figure \ref{fig:Streamline_7}.} Labels `L' and `VNTSA' indicate that initial conditions starting at the arrows evolve to laminar or VNTSA states, respectively. (b) Bifurcation diagram of the $L_{\infty}$-norm of $\hat{\alpha}_{xx}$ with $\Wi$ for VNTSA and EIT. Points (a)-(f) correspond to structures shown later \AS{in Figure \ref{fig:Structures_Wi40}}. The label `EIT' indicates that initial conditions evolve to EIT. \sout{Dashed curves are hypothesized unstable solution branches.} \AS{\sout{R1b - }In both plots, dashed lines correspond to the unstable solution branches obtained from edge tracking.} \sout{The asterisks on the plots indicate the continuation of the branch from one plot to the other.}}
		\label{fig:BFD}	     
	\end{center}
\end{figure}

{At the parameters chosen, the solution branch originating in the self-sustained Newtonian TS wave bifurcates to a periodic orbit at $\Wi \approx 3.5$ (cf.~\cite{Lee:2017fb}) before turning back into a traveling wave and losing existence beyond $\Wi = 3.875$, evidently in a saddle-node bifurcation yielding a lower branch TS wave solution that becomes the Newtonian solution as $\Wi\rightarrow 0$. Consistent with a saddle-node bifurcation, if the solution at $\Wi=3.875$ is used as an initial condition for a simulation at slightly higher $\Wi$, the flow laminarizes. This bifurcation scenario is shown on Fig.~\ref{fig:BFD}a in terms of average wall shear rate vs.~$\Wi$. \sout{Because we have not directly computed the lower branch, it is indicated as a dashed curve.} \AS{\sout{R1b - }The unstable lower branch (dashed blue) was found using edge tracking (\cite{Zammert:2014eq}) between NNTSA and laminar solutions at a given $\Wi$. A bisection technique was used to arrive at arbitrarily close initial conditions that are on either side of the edge. DNS trajectories starting from such points stay on the edge for a while before diverging to NNTSA and laminar.}}

As shown in \cite{Shekar:2019hq}, if $\Wi$ is large, sufficiently energetic initial conditions lead to 2D EIT. \sout{For comparison with the structure on the NNTSA branch, Figure \ref{fig:Streamline_7}(b) shows a snapshot of $\hat{\alpha}_{xx}$ on the EIT branch at $\Wi=15$.}  Fig.~\ref{fig:BFD}(a) also shows the mean wall shear rate for the EIT solution branch, which loses existence at finite amplitude when $\Wi\lesssim 13$. The bifurcation underlying this transition is presumably also of saddle-node form.

The central observation of the present paper arises from considering what happens just below the onset of the EIT regime at $\Wi\approx 13$. We do this by using a velocity and stress field from EIT at $\Wi  = 13$ as an initial condition for a run at $\Wi = 12$. This initial condition persists as a slowly decaying form of EIT for hundreds of time units, consistent with behavior just beyond a saddle-node bifurcation. As time increases further, the structure continues to decay,
 but does not ultimately reach the laminar state. Instead, 
 it evolves to a nontrivial attractor state that is very nearly a traveling wave, and in particular strongly resembles the linear TS mode at these parameters. We call this new state the viscoelastic nonlinear Tollmien-Schlichting attractor (VNTSA).
 
\subsection{Viscoelastic nonlinear Tollmien-Schlichting attractor\label{sec:lba}}

%

\begin{figure}
    \begin{center}
	\includegraphics[scale=0.28]{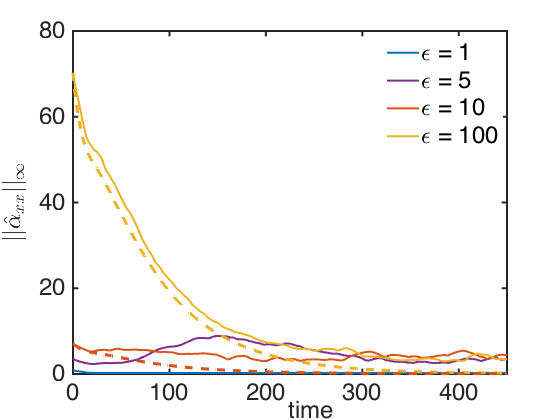}
	\caption[]{Time evolution of the $L_{\infty}$-norm of $\hat{\alpha}_{xx}$  for $Re = 3000, \Wi = 13, L_x=5$, starting from an initial condition of laminar state + $\epsilon\,\times\,$TS-mode. Dashed lines correspond to linearized runs starting from the same initial conditions for $\epsilon = 10$ and $100$. 
	}
 	\label{fig:L2_Cxxp_vs_time_Wi13}
    \end{center}
\end{figure}
\begin{figure}
	\begin{center}
		\begin{subfigure}{0.45\textwidth}
			\includegraphics[scale=0.38]{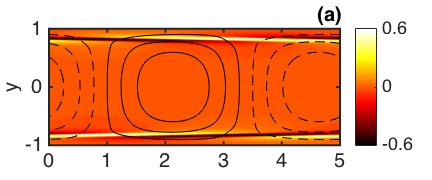} 
			\label{fig:TS_mode}
		\end{subfigure}
		\begin{subfigure}{0.45\textwidth}
	     \includegraphics[scale=0.38]{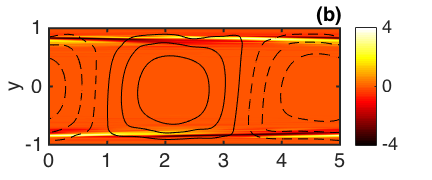} 
      	\label{fig:Combined_LBA}
       \end{subfigure}
		\begin{subfigure}{0.45\textwidth}
	\includegraphics[scale=0.38]{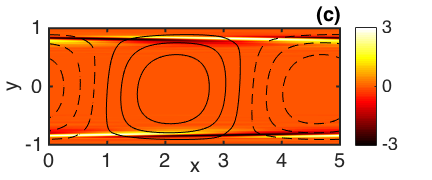} 
	\label{fig:LBA_Proj1}
		\end{subfigure} 
	\begin{subfigure}{0.45\textwidth}
		\includegraphics[scale=0.38]{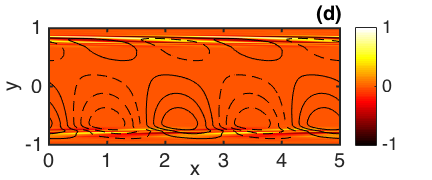} 
		\label{fig:LBA_Proj2}
	\end{subfigure}
\caption[]{(a) Structure of the linear TS mode at $Re = 3000, \Wi = 13, L_x=5$. Magnitude of the eigenmode is arbitrary and values shown here correspond to $\epsilon = 1$. (b): Snapshot of the fluctuation structure of the VNTSA at $Re = 3000, \Wi = 13$. (c) and (d): The $k L_x/2\pi  = 1, 2$ components respectively, of the snapshot shown in (b). Shown are contour lines of $\hat{v}$ superimposed on color contours of $\hat{\alpha}_{xx}$. }
\label{fig:LBA}	     
\end{center}
\end{figure}

{To elaborate on the relationship between the VNTSA and the linear TS mode, we \sout{now} describe \sout{2D simulation} results at $Re = 3000, \Wi = 13, L_x = 5$, i.e. close to the point where the 2D EIT branch first comes into existence as shown in the bifurcation diagram (Figure \ref{fig:BFD}a). EIT and the VNTSA are coexisting attractors at these parameter values.  }
\ASrevise{Figure \ref{fig:L2_Cxxp_vs_time_Wi13} shows the evolution of the $L_{\infty}$ norm of $\hat{\alpha}_{xx}$ starting from an initial condition consisting of the laminar state plus some amplitude $\epsilon$ of the linear TS mode for this parameter set. This mode, with  $\epsilon=1$, is shown in Figure \ref{fig:LBA}a. The structure of the velocity field is virtually unchanged from the Newtonian case and the polymer conformations are strongly localized to the critical layer positions $y = \pm 0.82$.  Sufficiently small perturbations, e.g.~$\epsilon=1$, decay to the laminar state, as they must since that state is linearly stable. However, larger perturbations $\epsilon=\AS{5}, 10$ and 100, where nonlinear mechanisms play a role, $\hat{\alpha}_{xx}$ settles to a finite value corresponding to the VNTSA. \AS{\sout{R2f - }The initial condition $\epsilon = 5$ that starts from below the VNTSA in $||\hat{\alpha}_{xx}||_{\infty}$ shows an initial growth phase before saturating onto the VNTSA, whereas    $\epsilon = 10$ and 100 relax onto the VNTSA from above.}} 

For comparison, the dashed lines on Figure \ref{fig:L2_Cxxp_vs_time_Wi13} show the linearized evolution starting from the same initial conditions; these all decay to laminar, illustrating the role of nonlinearity in the transition to the VNTSA. This state is robust: initial perturbation amplitudes over a wide range will evolve to it. However, initial conditions with very large magnitudes (e.g.~$\epsilon$ = 6000) evolve to EIT:  as noted above, both EIT and VNTSA are attractors at the chosen parameters (as is the laminar state).

\sout{\MDG{elaborate on what these perturbations look like - Done}}\AS{\sout{R3d - }We have also used initial conditions of the laminar state plus velocity perturbations somewhat similar to the ones used by \cite{Page:2015es}. These perturbations satisfied incompressibility and were sinusoidal in nature with streamwise and wall normal periods equal to the domain size. When appropriate magnitudes are used,  transient growth of polymer stress followed by a decay to the VNTSA is observed.}

\AS{\sout{R3b - }In 3D channel flow simulations of Oldroyd-B fluids at $\Rey = 3000$, $\Wi = 4$ to 6, $\beta = 0.9$, \cite{Min:2003id} observe a transient ($\sim 500$ time units) before an abrupt jump to the fully developed state. These observations were made \sout{in the ``large drag reduction" regime}starting from turbulent Newtonian initial conditions. In the present work, we used an average of about 2500 TU of data, and in some cases more than 5000 TU to calculate the statistics, and have observed no similar transition in any simulations of VNTSA in the parameter space considered here. }


Figure \ref{fig:LBA}b is a snapshot showing the typical fluctuation structure of the VNTSA at $\Wi=13$. The streamwise conformation $\hat{\alpha}_{xx}$ has tilted sheets highly localized near $y = \pm 0.82$ and contours of wall normal velocity $\hat{v}$ span the entire channel. This  structure bears strong resemblance to the TS mode shown in Figure \ref{fig:LBA}a. The VNTSA is thus a weakly nonlinear self-sustaining state whose primary structure is the viscoelastic TS mode. We elaborate in the following section on the linear TS mode and its connections to the VNTSA.


In the VNTSA state, the velocity fluctuations are very weak, and the mean wall shear rate displays a very small change from laminar. This can be understood on the grounds that changes of the mean wall shear rate correspond to fluctuations with $k_x=0$, which arise only due to nonlinear interactions. Since the primary velocity structure is very weak, the nonlinear effects will be even weaker.
\ASrevise{To illustrate nonlinear effects, Figures \ref{fig:LBA}c and  \ref{fig:LBA}d, respectively, show the $k L_x/2\pi  = 1, 2$, spatial Fourier components of the snapshot shown in \ref{fig:LBA}b. Figure \ref{fig:LBA}c closely resembles the TS mode, with a slight symmetry-breaking across the centerline $y=0$. The structure at $k L_x/2\pi  = 2$ also displays polymer stress fluctuations localized around the critical layer position, an observation that also holds for higher wavenumbers. 
}

\ASrevise{Having established the structure of the flow on the VNTSA branch, we now illustrate the bifurcation scenario of this solution branch by continuing in $\Wi$. The VNTSA branch loses existence at finite amplitude (i.e.~in a saddle-node bifurcation) for $\Wi\lesssim 6$, as we have confirmed both by using the $\Wi=6$ solution as an initial condition for simulations at lower $\Wi$ and by running simulations starting from the laminar state perturbed by the TS mode with small $\epsilon$. For $Wi<6$ all these initial conditions decay to laminar.
 On increasing $\Wi$, the VNTSA branch seems to lose existence \ASrevise{beyond $\Wi \approx 49$}, and initial conditions that land on the VNTSA for $\Wi = 49$ evolve to EIT at $\Wi=50$. \sout{These observations suggest that the VNTSA turns around and forms an unstable branch that joins up with the unstable lower branch of EIT.}\sout{\AS{Move to end of next para}} \sout{Due to the small amplitude of the VNTSA branch, the bifurcation scenario associated with it is shown separately in Figure \ref{fig:BFD}b,} \AS{Due to its weak nature in relation to EIT, the bifurcation scenario associated with the two solution branches is shown on Figure \ref{fig:BFD}b in a log scale} using the $L_{\infty}$ norm of $\hat{\alpha}_{xx}$ as the amplitude measure. \sout{The hypothesized unstable branch connecting VNTSA and EIT is shown schematically with dashed lines on the bifurcation diagrams}}\AS{The EIT solutions in Figure \ref{fig:BFD}a are depicted again in Figure \ref{fig:BFD}b as the upper stable branch (solid red) with this measure.}
 
 \AS{\sout{R1b - }Since the VNTSA and EIT are both stable states, we were also able to perform edge tracking to find unstable solutions intermediate between these two states. Five $\Wi$ values (13, 15, 20, 40 and 45) were studied. \AS{Repeated bisections were performed until trajectories stayed on the edge for an average of 300 TU.} The red-dashed line on Figures \ref{fig:BFD}a and b indicates solutions (all time-dependent) on this intermediate branch.  The magnitude of fluctuations along this branch monotonically decreases on increasing $\Wi$, displaying values close to EIT at $\Wi = 13$ and values to close to VNTSA at $\Wi = 45$. Furthermore, this  branch also displays fluctuations that resemble the viscoelastic TS mode, as illustrated below. 
 A more detailed link between VNTSA and EIT might be established through numerical continuations of underlying traveling wave solutions; this is a topic of future endeavors.}
 
 \AS{\sout{R1d - }To complete this discussion, we note that the bifurcation scenario we observe implies the existence of an edge between the VNTSA and the laminar state, which could in principle also be found using edge tracking. However, the weak nature of the VNTSA implies that this edge would be even weaker, thus making this a challenging task.}

\begin{figure}
	\begin{center}
		\begin{subfigure}{0.45\textwidth}
			\includegraphics[scale=0.38]{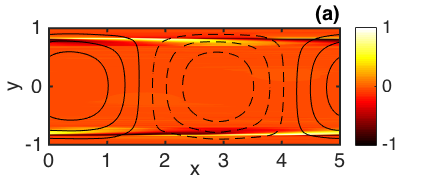} 
			\label{fig:Combined_1500_Wi8}
		\end{subfigure}
		\begin{subfigure}{0.45\textwidth}
			\includegraphics[scale=0.38]{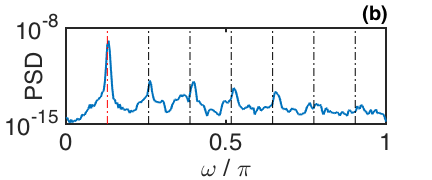} 
			\label{fig:Pow_Den_vp_0p825_Wi8_Welchs}
		\end{subfigure}
		\begin{subfigure}{0.45\textwidth}
			\includegraphics[scale=0.38]{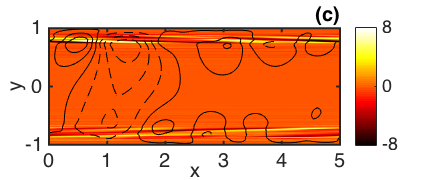} 
			\label{fig:Combined_2300_Wi20}
		\end{subfigure}
		\begin{subfigure}{0.45\textwidth}
			\includegraphics[scale=0.38]{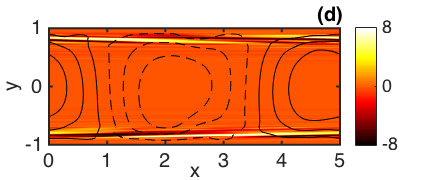} 
			\label{fig:Combined_3100_Wi20}
		\end{subfigure}
		\caption[]{(a): Fluctuation structure of the VNTSA and (b) power spectral density (PSD) of $v$ at position $(0,0.825)$ at $\Rey = 3000$, $\Wi = 8$. Frequencies corresponding to the TS mode (red-dashed) and its higher harmonics (black-dashed) are also shown. (c) and (d): snapshots of the VNTSA structure for $\Rey = 3000$, $\Wi = 20$. Contour plots follow the same format as in Figure \ref{fig:LBA}.}
		\label{fig:LBA_Structure_vs_Wi}	     
	\end{center}
\end{figure}

\ASrevise{Figure \ref{fig:LBA_Structure_vs_Wi}a shows the fluctuation structure of the VNTSA at $\Wi = 8$, close to the point where it first comes into existence. The structure closely resembles the TS mode and does not change appreciably with time. The flow is almost a pure nonlinear traveling wave with some weak non-periodicity, as indicated in the power density plot of the wall normal velocity at position $(0,0.825)$ shown in Figure \ref{fig:LBA_Structure_vs_Wi}b. The spectrum is mainly composed of the dominant TS mode frequency and its higher harmonics. The dynamics and structures get more complicated as $\Wi$ increases. Figure \ref{fig:LBA_Structure_vs_Wi}c shows a typical snapshot at $\Wi = 20$, which clearly is more complex than a TS mode.  However, at this $\Wi$, the VNTSA still intermittently displays clear TS-like structures such as the snapshot in Figure \ref{fig:LBA_Structure_vs_Wi}d.}

\begin{figure}
	\begin{center}
        \begin{subfigure}{0.45\textwidth}
			\includegraphics[scale=0.38]{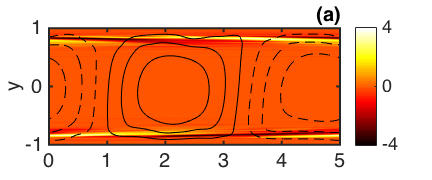} 
			\label{fig:VNTSA_Wi13}
		\end{subfigure}
		\begin{subfigure}{0.45\textwidth}
			\includegraphics[scale=0.38]{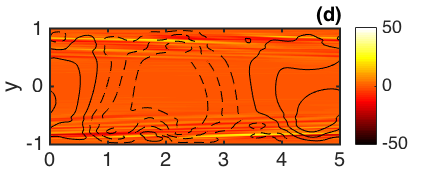} 
			\label{fig:LBA_Wi40}
		\end{subfigure}
		\begin{subfigure}{0.45\textwidth}
			\includegraphics[scale=0.38]{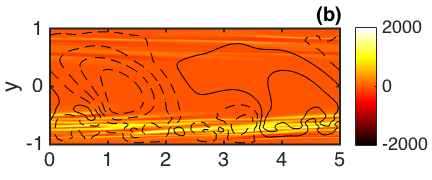} 
			\label{fig:Edge_Wi13_2}
		\end{subfigure}
		\begin{subfigure}{0.45\textwidth}
			\includegraphics[scale=0.38]{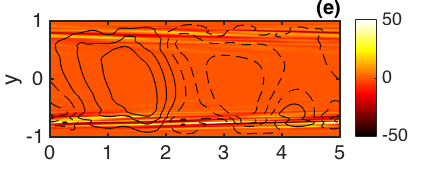} 
			\label{fig:EIT_Wi40}
		\end{subfigure}
		\begin{subfigure}{0.45\textwidth}
			\includegraphics[scale=0.38]{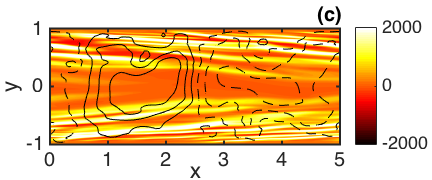} 
			\label{fig:EIT_Wi15}
		\end{subfigure}
		\begin{subfigure}{0.45\textwidth}
			\includegraphics[scale=0.38]{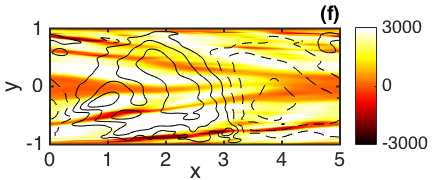} 
			\label{fig:EIT_Wi40}
		\end{subfigure}
		\caption[]{\AS{\sout{R1c, R2d - }(a), (b) and (c): Snapshots of VNTSA, intermediate branch and EIT respectively at $\Wi = 13$. (d), (e) and (f): Snapshots of the same solution branches at $\Wi = 45$. Deviations from the laminar base state shown as earlier. Color scales are intentionally centered about 0 for comparison purposes.}}
		\label{fig:Structures_Wi40}	     
	\end{center}
\end{figure}

{We now turn to the flow structures at various positions on the bifurcation diagram, Figure \ref{fig:BFD}b.} \AS{\sout{R1c, R2d - }Figures \ref{fig:Structures_Wi40}a, b and c are representative snapshots of VNTSA, the intermediate branch and EIT respectively, at $\Wi = 13$ i.e. near the loss of existence of EIT. As detailed in the description of Figure \ref{fig:LBA} and shown again in Figure \ref{fig:Structures_Wi40}a, VNTSA exhibits a weak structure that strongly resembles the viscoelastic TS mode. At this low $\Wi$, the intermediate branch (Figure \ref{fig:Structures_Wi40}b) displays a much stronger fluctuation structure, on the same order of magnitude as the structures seen at EIT (Figure \ref{fig:Structures_Wi40}c). Moreover, the intermediate branch exhibits overlapping sheets of polymer stretch(especially on the bottom side of the channel at the time instant shown) that strongly resemble those seen at EIT. 
}

\AS{\sout{Moving along the branches, }Figures \ref{fig:Structures_Wi40}d, e and f show the structures at $\Wi = 45$, close to the point where the intermediate branch seems to turns around to merge with VNTSA. The fluctuations on the intermediate branch (Figure \ref{fig:Structures_Wi40}e) decrease in magnitude on increasing $\Wi$, and at $\Wi=45$ become comparable to those on VNTSA branch as shown in Figure \ref{fig:Structures_Wi40}d. Furthermore, structure on this branch goes from displaying overlapping sheets of polymer stretch at low $\Wi$ to localized striations at high $\Wi$ similar to those of VNTSA. In sharp contrast to the localized structures just described, EIT (Figure \ref{fig:Structures_Wi40}f) continues to display sheets of polymer stretch which get stronger on increasing $\Wi$. All three solution branches continue to intermittently display wall normal velocity fluctuations that resemble the TS mode. \sout{Taken together, these observations are consistent with the intermediate branch being the unstable branch joining VNTSA and EIT.}}

\subsection{Linear analyses}

In this section we elaborate on the linearized problem and its connection to the attractors described above using linear stability and resolvent analyses. The spectrum corresponding to disturbances with wavelengths equal to the DNS box size, i.e. $k=2\pi/5$, has a least-stable eigenvalue at $c \approx 0.32-0.010i$, and the associated eigenfunction is the viscoelastic extension of the TS mode. {For low values of $\Wi$, the mode is less stable than its Newtonian counterpart, while for $\Wi\gtrsim 2$, it becomes more stable with increasing elasticity}; this non-monotonic behavior has been reported by \cite{zhang2013}, who attribute it to viscoelastic modification of the phase difference between $u$ and $v$. \AS{Nevertheless,} over the range of $\Wi$ considered here, the eigenvalue varies by less than 1\% of the Newtonian value. {The linear stability of the laminar state in this range of $\Wi$, and the very weak dependence of $c$ on $\Wi$ continues up to at least $\Rey=6000$,}
confirming the observations in Section~\ref{sec:lba} that finite amplitude disturbances are required to trigger transition to EIT or the VNTSA. However, linear instabilities not related to the TS mode have been found in other regions of parameter space \citep{garg2018,chaudhary2019}, implying the possibility of different attractor families in those regions.

\begin{figure}
    \centering
    \includegraphics[width=\linewidth]{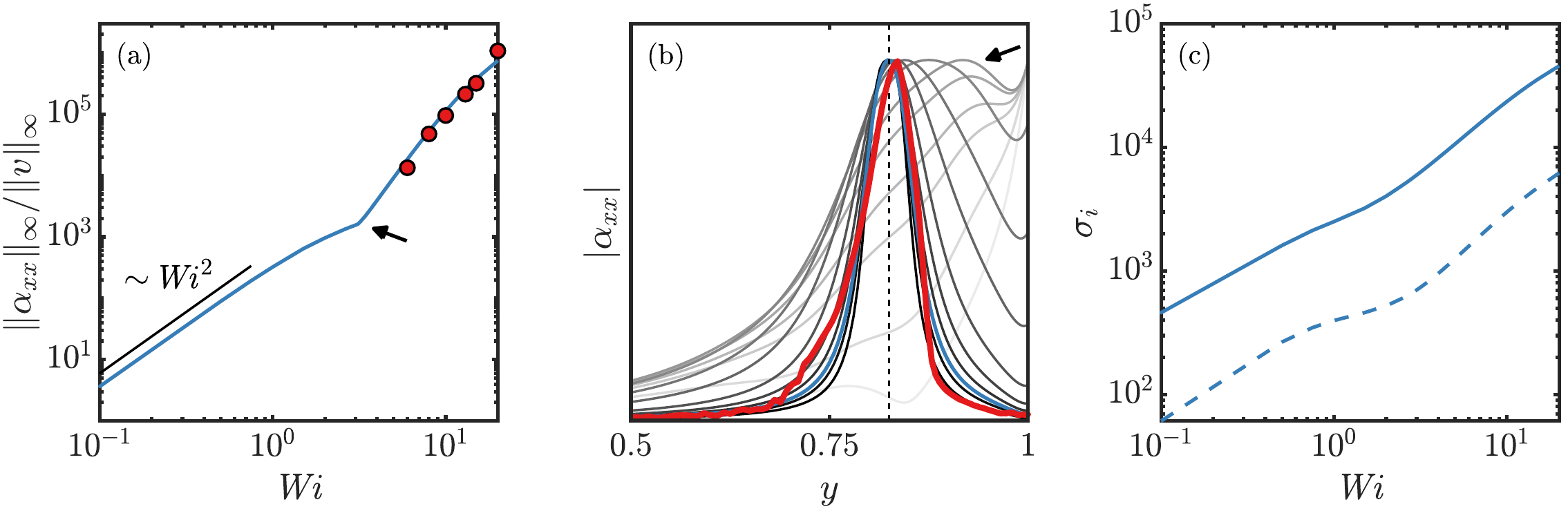}
	\caption{
	(a) Solid blue line: ratio of peak amplitudes of $\hat{\alpha}_{xx}$ and $\hat{v}$ for the linear TS mode as a function of $\Wi$; circles: amplitude ratio for the VNTSA. (b) Magnitude of $\hat{\alpha}_{xx}$ for the linear TS mode  
	for several values of $\Wi$ in the range $[1,\,20]$. Darker lines indicate higher values of $\Wi$. The thick red line shows the averaged magnitude of $\hat{\alpha}_{xx}$ from the VNTSA for $\Wi=13$. For comparison, the linear TS mode profile for the same $\Wi$ is shown in  blue, and the vertical dashed line marks the critical layer location $y_c = 0.825$. The arrow indicates the value of $\Wi$ corresponding to the arrow in (a). (c) First two singular values of the resolvent operator for $k$ and $c$ corresponding to the linear TS mode. 
	}
\label{fig:TSW}
\end{figure}

A measure of the relative importance of the conformation tensor and velocity disturbances is the ratio of the peak amplitudes of $\hat{\alpha}_{xx}$, (the largest component of the conformation tensor), and $\hat{v}$. This ratio is shown in Figure~\ref{fig:TSW}a. 
Two distinct regimes are apparent, with the transition between the two occurring at $\Wi\approx 3.1$. The low $\Wi$ regime scales as $\Wi^2$, which is the same scaling as in linear shear flow. The amplitude ratio above the change in slope does not exhibit power law scaling. 
The change in slope at $\Wi\approx 3.1$, can be understood by examining the $\hat{\alpha}_{xx}$ mode shapes, the magnitudes of which are plotted in Figure~\ref{fig:TSW}b for several values of $\Wi$ in the range shown in Figure~\ref{fig:TSW}a. For small $\Wi$, the disturbance is largest at the wall and decays rapidly away from it. Therefore, the $\Wi^2$ scaling in this regime can be explained by the fact that the leading-order approximation of the base flow very near the wall is simple shear. As $\Wi$ increases, this value decreases, while a new local maximum emerges and grows, becoming the global maximum just above $\Wi=3$; the arrow in the figure indicates the profile where this occurs. Upon further increase in $\Wi$, the maximum gradually shifts away from the wall, and the modes become increasingly localized around 
the location of the critical layer $y_c$, at which the real part of the wavespeed equals the base flow velocity. The critical layer for $\Wi=13$ is indicated by the vertical dashed line. This suggests that a critical layer mechanism is responsible for the change in 
scaling at large $\Wi$, though at present we do not understand the specific origin of this result. Interestingly, the $\Wi$ at which the VNTSA comes into existence is only slightly larger than that at which the transition to critical layer scaling occurs.

Also shown in Figure~\ref{fig:TSW}a is the amplitude ratio computed from the VNTSA for several values of $\Wi$. Excellent agreement between the linear and nonlinear results quantitatively reinforces the TS-mode-like nature of the VNTSA. Additionally, the profile of $\vert\hat{\alpha}_{xx}\vert$, averaged in the streamwise direction and over many snapshots, for the VNTSA at $\Wi=13$ is shown by the thick red line in Figure~\ref{fig:TSW}b, and the blue line highlights the linear mode for the same $\Wi$. The VNTSA profile exhibits the same localization, and the location of the peak value is in close agreement with the critical layer location.

Figure~\ref{fig:TSW}c shows the first two singular values of the resolvent operator for $k$ and $c$ corresponding to the linear TS mode. \cite{Shekar:2019hq} showed that such modes are the most-amplified 2D disturbances in this parameter regime and that the leading response mode is nearly identical to the TS eigenmode; for this reason the resolvent modes are not plotted separately. The substantial increase in the leading singular value with $\Wi$ indicates that this amplification becomes much stronger with increasing elasticity, and consequently that considerably smaller disturbances may be sufficient to trigger self-sustaining nonlinear mechanisms. Further, the symmetry of the flow geometry about $y=0$ means that resolvent modes typically come in pairs having similar amplification, with one mode having a symmetric $\hat{v}$ response and the other having an antisymmetric $\hat{v}$ response. However, the growing separation between the first and second singular values with increasing $\Wi$ indicates that this pairing is broken by elasticity, and that the symmetry exhibited by the TS mode is preferred in terms of linear amplification.

\subsection{Broader context: flow geometry and dimensionality}

The results described above are limited to two-dimensional channel flow. Here we describe how they may be viewed in the broader context of turbulent drag reduction, first with regard to how they may relate to the pipe flow geometry and then in the full context of three-dimensional turbulence. 
 
As noted in the Introduction, elastoinertial turbulence with very similar features has been observed in both channel and pipe flows. A natural question, then, is to what extent the above channel flow results are relevant to the pipe flow case. 
While it is true of course that there is no linear instability of Newtonian pipe flow, the general structures of the linear stability problem for the pipe and channel are very similar. The A, P, and S (wall, center, strongly decaying) families of modes for the channel flow problem also arise in the pipe, as illustrated in Figs. 4.18 and 4.19 of \cite{DrazinReid}. (In channel flow, one of the wall modes, the TS mode, goes unstable.) 
 Additionally, critical layers are a strong source of linear amplification in pipe flow, as they are in channel flow \citep{Mckeon:2010ep}. 
 
 
\MDGrevise{Turning to {nonlinear} behavior in Newtonian pipe flow, an important difference from channel flow is the apparent absence of subcritical traveling wave solutions that are analogous to the nonlinear Tollmien-Schlichting waves \citep{Patera:1981ez}.   That is, in pipe flow there is no nonlinear solution branch that corresponds to the NNTSA described above.  Given this observation, one might wonder whether the results presented here for viscoelastic channel flow are relevant for pipe flow.}  

\MDGrevise{To address this issue, we make the following remarks.  The key observation of the present work is that the TS mode is {nonlinearly} excited  \emph{by viscoelasticity}. The resulting solution branch, the VNTSA, is not connected, at least in this part of parameter space under consideration, to the Newtonian branch, indicating that the nonlinear mechanism that sustains it is distinct from the nonlinear mechanisms sustaining the Newtonian branch. So the absence of a \emph{Newtonian} mechanism for nonlinear sustainment of Tollmien-Schlichting-like traveling waves does not imply the absence of a \emph{viscoelastic} mechanism.  Furthermore, as noted in the Introduction, pipe flow simulations of EIT \citep{lopez2019dynamics} display  essentially two-dimensional velocity fluctuations localized near the wall that are similar to those reported in channel flow. These are precisely what would be expected in pipe flow based on the observations we report here for channel flow, i.e.  a critical layer mechanism associated with excitation of a Tollmien-Schlichting-like mode.} \AScomment{At the same time, in more strongly viscoelastic regimes, \cite{garg2018} and \cite{Chaudhary:2020wd} have found a center mode instability for pipe flow; for the Oldroyd-B model with $\Rey=3500, \beta=0.9$, instability occurs when $176.9 < \Wi < 4783.6$. A related instability might be present in the channel flow problem. These results open up the possibility that other states unrelated to the nonlinear excitation of a wall mode may also play a role at EIT in both channels and pipes, especially at high $\Wi$.}



We now turn to the topic how the present results are related to the fully three-dimensional context, beginning with a brief overview of Newtonian near-wall turbulence. 

Newtonian turbulence is of course, strongly three-dimensional, with the dominant near-wall structure comprised of coherent wavy streamwise vortices.
In all of the canonical wall-bounded shear flow geometries (pipe flow, channel flow, plane shear (Couette) flow), families of three-dimensional nonlinear traveling wave solutions to the Navier-Stokes equations (NSE) have been discovered \citep{Waleffe:1998wk,Waleffe:2001wu,Waleffe:2003hh,Wang:2007ii,Hof:2004ug,Eckhardt:2007ka,Eckhardt:2008gd,Duguet:2008bs,Wedin:2004ey}.   These solutions are often denoted ``exact coherent states'' (ECS), and their predominant structure is very similar to that observed in wall turbulence: a mean shear and wavy streamwise vortices. (In particular, they bear no structural resemblance to Tollmien-Schlichting waves and do not arise from a linear instability of the laminar state.) Related, but more complex states have been found as well, that are not pure traveling waves but rather ``relative periodic orbits'' that are time-periodic modulo a phase shift in one of the translation-invariant spatial directions \citep{Duguet:2008ev}. In minimal domains at Reynolds numbers near transition, the turbulent dynamics have been found to be organized, at least in part, around these solutions (see, e.g.~\cite{Gibson:2008eca,Kawahara:2012iu,Park:2014wt}).

It has long been known that one of the effects of viscoelasticity on wall turbulence is to weaken and broaden the near-wall streamwise vortices \citep{Kim:2007dq,White:2008hs}. A number of studies have addressed this observation by investigating the effect of viscoelasticity on ECS, which as noted, capture this streamwise vortex structure \citep{Stone:2002dj,Stone:2003gq,Stone:2004jk,Li:2005vl,Li:2006gk,Li:2007ii}. Indeed, the effect of viscoelasticity is to weaken these structures; the polymer stresses directly counteract the streamwise vortices.

In particular, \cite{Li:2007ii} studied the bifurcation scenario for a particular family of channel flow ECS (\cite{Waleffe:1998wk}) in a parameter regime very close to that considered here. At $\Rey=1500$, this ECS family is sufficiently weakened by viscoelasticity to lose existence at $\Wi\approx 16$, somewhat above the value $\Wi\approx 10$ beyond which Newtonian turbulence cannot self-sustain in the DNS study of \cite{Shekar:2019hq}. (This discrepancy is consistent with what we know about transition in the Newtonian case: channel flow turbulence is self-sustaining above $\Rey\approx 1000$, while ECS can exist in that case down to $\Rey\approx 660$ \citep{Shekar:2018gg}.) Extrapolating slightly from the results of \cite{Li:2007ii}, one can estimate that at $\Rey=3000$, this ECS family loses existence at $\Wi\approx 25$. 

Given that, in the present 2D work with $\Rey=3000$, EIT is found at $\Wi\gtrsim 13$ and the VNTSA above at $\Wi\gtrsim 6$, there would appear to be a regime $6\lesssim \Wi\lesssim 25$ in which both 2D and 3D structures may exist and interact. Furthermore, existing results on the effect of viscoelasticity on ECS are limited to one ECS family, and there certainly may be others that can persist to higher $\Wi$. Consistent with this analysis, a number of studies have reported near-wall EIT-like spanwise oriented structures, with 3D quasistreamwise vortices further away from the wall \citep{Dubief:2013hh, Choueiri:2018it,  pereira2019beyond, pereira2019common}. 
How the 2D and 3D structures interact is an important topic for future work.

\section{Conclusion}

This study focuses on two-dimensional plane channel flow of a very dilute polymer solution at $\Rey=3000$. At sufficiently high $\Wi$, elastoinertial turbulence is observed in this parameter regime, and the focus of the present work is to make progress toward understanding the structures and mechanisms underlying the dynamics in this regime. 
We report here the existence of a new attractor that is based on the viscoelastic linear Tollmien-Schlichting mode and is nonlinearly sustained by viscoelastic stresses. We denote this as the viscoelastic nonlinear Tollmien-Schlichting attractor (VNTSA).
At the parameters considered here, this solution branch is not connected to the Newtonian branch of nonlinear self-sustained Tollmien-Schlichting waves; it would be interesting to learn whether they become connected at higher $\Rey$. 
In a domain of dimensionless length $5$, 
this solution comes into existence at finite but very small amplitude when $\Wi\gtrsim 6$, increasing in amplitude until $\Wi\approx 49$ where it loses existence again. At higher $\Wi$, initial conditions corresponding to this solution branch at lower $\Wi$ evolve into elastoinertial turbulence. 
In general, we do not find pure nonlinear traveling waves, but until $\Wi$ is large, the nonperiodic fluctuations are very small. 
The connection of the VNTSA to the linear TS mode is established via their strong structural similarities, including a quantitive agreement between the relative magnitudes of the velocity and stress fluctuations.
The value of $\Wi$ at which the VNTSA comes into existence is close to where the relative amplitude of the stress and velocity fluctuations for the linear TS mode undergoes a change in scaling. Above this transition the stress fluctuations become highly localized at the position of the critical layer. 

Taken together, these results suggest that, at least in the parameter range considered here, the bypass transition leading to EIT is mediated by nonlinear amplification and self-sustenance of perturbations that excite the Tollmien-Schlichting mode. Gaining an understanding of the mechanism underlying this phenomenon will shed light on the origin of elastoinertial turbulence. 

\section*{Acknowledgments}
This work was supported by (UW) NSF  CBET-1510291, AFOSR  FA9550-18-1-0174, and ONR N00014-18-1-2865, and (Caltech) ONR N00014-17-1-3022.

Declaration of interests: none.

\bibliographystyle{jfm}
\bibliography{turbulenceMDG,papersMDG,BJM_RMM_papers,pipestability}

\end{document}